\documentclass[12pt,preprint]{aastex}

\newcommand{\ax}{{AX~J08494+4454}}
\newcommand{\ergs}{ergs cm$^{-2}$ s$^{-1}$}

\newcommand{\asca}{{\it ASCA}}

\newcommand{\oiii}{{[O~III]$\lambda$5007}}
\newcommand{\sii}{{[S~II]$\lambda\lambda$6718,6733}}

\newcommand{\nii}{{[N~II]$\lambda\lambda$6548,6583}}
\newcommand{\niia}{{[N~II]$\lambda$6548}}
\newcommand{\niib}{{[N~II]$\lambda$6583}}
\slugcomment{Accepted for publication in ApJ}

\shorttitle{Absorbed QSO}
\shortauthors{Akiyama, Ueda, and Ohta}

\begin{document}

\title{A candidate of a type-2 QSO at $z=0.9$:
Large X-ray absorption with a
strong broad-H$\alpha$ emission line \altaffilmark{1}}

\author{Masayuki Akiyama}
\affil{Subaru Telescope, National Astronomical Observatory of Japan,
Hilo, HI, 96720}
\email{akiyama@subaru.naoj.org}

\author{Yoshihiro Ueda}
\affil{Institute of Space and Astronautical Science, Sagamihara,
Kanagawa, 229-8510, Japan}
\email{ueda@astro.isas.ac.jp}

\and

\author{Kouji Ohta}
\affil{Department of Astronomy,
Kyoto University, Kyoto, 606-8502, Japan}
\email{ohta@kusastro.kyoto-u.ac.jp}

\altaffiltext{1}{
Based on data collected at Subaru Telescope,
which is operated by the National Astronomical Observatory of Japan,
and University of Hawaii 2.2m Telescope, HI.}

\begin{abstract}
Deep hard X-ray and near infrared observations of a
type-2 radio-quiet QSO candidate at $z=0.9$, \ax, are reported.
The 0.5--10~keV {\it Chandra X-ray Observatory} spectrum of \ax\
is hard, and is explained well with a power-law continuum absorbed
by a hydrogen column density of $(2.3\pm1.1) \times 10^{23}$ cm$^{-2}$.
The 2--10~keV luminosity of the object is estimated to be
$7.2^{+3.6}_{-2.0} \times 10^{44}$ erg s$^{-1}$, after correcting
the absorption, and reaches hard X-ray luminosities of QSOs.
The large X-ray absorption  and the
large intrinsic luminosity support the original identification of
\ax\ as a type-2 radio-quiet QSO.
Nevertheless, 
the deep Subaru/IRCS $J$-band spectroscopic
observation suggests the presence of a strong 
broad H$\alpha$ emission line from  \ax. If real, 
the broad H$\alpha$ emission line has a velocity width of
$9400\pm1000$ km s$^{-1}$, which corresponds to a typical
broad-Balmer line velocity width of a luminous QSO.
The existence of the strong broad H$\alpha$ line means that
the object is not a type-2 QSO, but a luminous cousin of
a Seyfert 1.9 galaxy in the source-frame optical spectrum.
The Balmer decrement of broad lines, and
the broad H$\alpha$ emission to the hard X-ray luminosity ratio
suggest that the nucleus is affected by dust extinction with
$A_V$ of $1-3$ mag in the optical wavelength.
Optical colors and faint optical magnitudes of \ax\
are consistent with the model.
The estimated amount of dust extinction is much smaller than that
expected from the X-ray column density ($A_V=130\pm60$ mag).
\end{abstract}


\section{Introduction}

"Type-2 QSO" is a missing population of AGNs which are
radio-quiet narrow-line QSOs, i.e., luminous cousins of
Seyfert 2 galaxies, or radio-quiet analogues of
high-redshift radio galaxies. So-called unified scheme of
Seyfert galaxies explains differences between Seyfert 1
and 2 galaxies with a difference of viewing angle to a
nucleus surrounded by a dusty torus system \citep{ant93}.
QSOs are thought to be luminous versions of Seyfert galaxies.
If the unified scheme holds for QSOs, then they should have
a similar nucleus and dusty torus system in their center.
Seyfert 2 galaxies outnumber Seyfert 1 galaxies in the
local universe \citep{huc92}, 
and if that ratio can be extrapolated to higher
luminosities, then we should expect many type-2 QSOs to be
detectable in the high-redshift universe. 
Strikingly, none have been convincingly detected to date.
The existence of absorbed high luminosity AGN has been
postulated to account for the hard X-ray spectrum of the 
cosmic X-ray background;
there should be
three times more absorbed narrow-line QSOs than
non-absorbed broad-line QSOs \citep{com95}.

Recently, deep hard X-ray surveys with imaging satellites detect
several candidates of type-2 QSOs in the high-redshift universe
(\citet{alm95}; \citet{oht96}, hereafter Paper I;
\citet{nor01}; \citet{daw01}).
These objects have hard X-ray luminosities as luminous as broad-line QSOs,
and their hard X-ray spectra suggest that their X-ray nuclei are
obscured. In the observed optical wavelength, they do not show
the broad components in permitted emission lines, like Seyfert 2 galaxies.
But this evidence is insufficient to definitely identify
these objects as luminous versions of Seyfert 2 galaxies,
because the characteristic point of Seyfert 2 galaxies
is that no broad H$\alpha$ emission line (above a certain level).
Therefore, in order to discuss similarities to Seyfert 2 galaxies,
we need to cover source-frame H$\alpha$ emission line.
It should be noted that majority of AGNs in the local
universe meets this criterion \citep{mai95}.
Actually, for the candidate of type-2 QSO found by
Almaini et al. (1995),
later observation in the near-infrared wavelength
revealed the presence of a strong broad H$\alpha$ emission line and the
object was reassigned as a type-1.9 QSO \citep{geo99}.
So far there is no hard-X-ray-selected type-2 QSO which is
confirmed to follow the strict definition of Seyfert 2 (see also
\citet{hal99}).
If there is no true type-2 QSO in the universe,
it is suggested that QSOs are not simple luminous
extension of Seyfert galaxies.

\ax\ is another candidate of a type-2 QSO at $z=0.9$.
It was found in the course of the optical identification of
{\it ASCA} deep survey in Lynx field (Paper I).
Based on the large hard X-ray luminosity,
hard X-ray spectrum, which reflects a heavy absorption
to the nucleus, and absence of broad H$\beta$ emission line,
the object is identified as a type-2 QSO (Paper I).
Later $J$-band spectroscopic observation targeted on the
redshifted H$\alpha$ emission line does not show the presence of
a strong broad H$\alpha$ emission line at all (\citet{nak00},
hereafter Paper II).

In this paper, results of
a deep hard X-ray observation with {\it Chandra X-ray Observatory} and
a deeper near-infrared observation of a redshifted
H$\alpha$ emission line with Subaru telescope of
\ax\ are reported.
The X-ray spectrum of the object establishes the large
absorption column density along the line of sight to the nucleus.
Although the previous near-infrared spectroscopy of \ax\
did not detect a significant broad H$\alpha$ emission line,
the deeper near-infrared spectrum reveals the probable existence of a very
broad H$\alpha$ emission line, whose strength suggests the
broad-line is not affected severely by dust absorption. We discuss the
discrepancy between source-frame optical
and X-ray absorption and no existence of "type-2 QSOs".
Throughout this paper,
we use $q_0=0.5$ and $H_0=50$ km s$^{-1}$ Mpc$^{-1}$.

\section{Observations and Results}

\subsection{X-ray Observation with {\it Chandra}}

\subsubsection{Observation}

\ax\ was observed with the {\it Chandra X-ray Observatory} in the
0.5--10 keV band on 2000 May 3 (ObsID=1708) and May 4 through 6
(ObsID=927) in the course of deep observations of the Lynx field
(object name = CL~0848.6+4453). \ax\ was located in the ACIS-I3 chip
at the row number of about 800. To mitigate the effect of degradation
by the Charge Transfer Inefficiency (CTI), we applied an improved
technique developed by \citet{tow00} for the the level 1 event
data. This improved the spectral resolution to 4.4\% (FWHM) at 3.4 keV
(corresponding to 6.4 keV at $z=0.8858$). Then the data were analyzed
in a standard way using the CIAO 2.1 software. We used only events of
\asca\ grade 0, 2, 3, 4, and 6. After excluding periods of the
background flares, we obtained a net exposure of 45 ksec for the May 3
observation and 119 ksec for the May 4--6 observation.

We detected \ax\ in the ACIS images as a point-like source from
the both data sets. The position determined by {\it Chandra} is
$\alpha = 8^{\rm h} 49^{\rm m} 27^{\rm s}.77$,
$\delta = +44^{\circ} 54^{\prime} 57.\!^{\prime\prime}6$ (J2000),
after correcting the aspect solution according to the method available
at the {\it Chandra} Science Center web page. The position error was
conservatively estimated to be 1 arcsec: we confirmed the accuracy by
cross correlating the position of a Tyco-catalog star and other APM
catalog sources located in the field of view of the ACIS-I arrays. The
X-ray source position is $0.\!^{\prime\prime}5$ east and
$0.\!^{\prime\prime}6$ south of the optical position, and the X-ray
and optical positions agree with each other within the
uncertainty. The accurate X-ray position clearly rejects the
possibility that the X-ray emission comes from a no-emission-line
object $2^{\prime\prime}$ west (see Paper I and II).

\subsubsection{Spectral Analysis}

For an extraction of the spectrum, we accumulated events around the
source position within a radius of 5 arcsec. The background spectrum
was taken from a nearby, bright-source free region in the same
chip. The background count rate was only 1\% to the source photons in
the 0.5--10 keV band and was not important. Examining the light curve
and spectrum, we found that both the flux level and spectral shape of
\ax\ were constant throughout the two observations within the
statistical error. We thus summed the spectra of the two data sets
with a total exposure of 164 ksec to obtain the best photon
statistics. For a spectral fit, we used an appropriate response
matrix file in which the above CTI correction is taken into account.
Considering possible calibration uncertainties at low energies, we
added a systematic error of 20\% to each spectral bin below 1.2
keV. Spectral analysis was performed using the XSPEC package.

\begin{figure}
\plotone{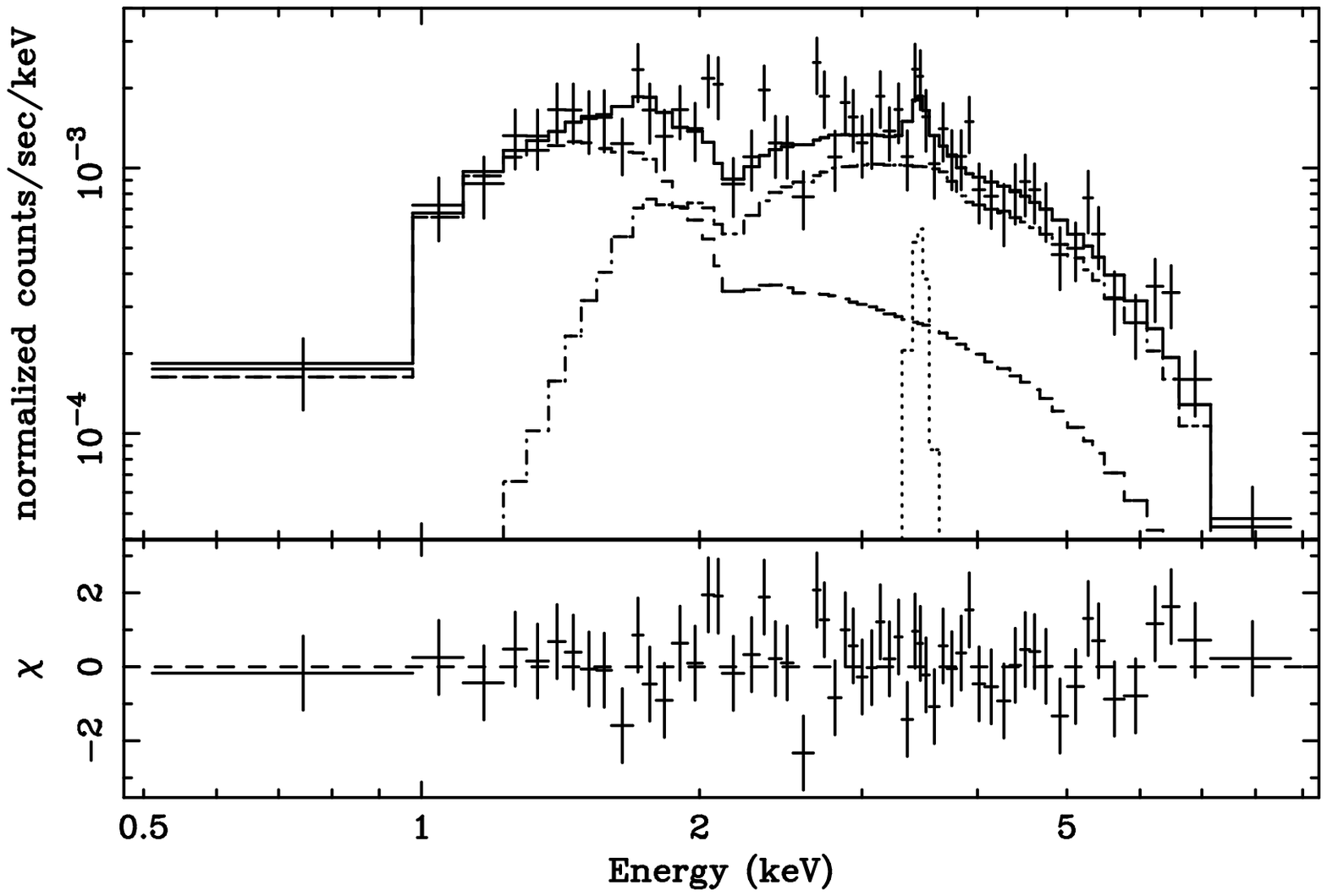}
\caption{
The {\it Chandra}-ACIS spectrum of AX~J08494+4454. The best-fit model
is plotted with solid line. The components of the model, 
the two power laws with different absorptions (dot-dashed and
dashed lines) and an iron K emission line (dotted line), 
are also plotted.
\label{chan_spec}}
\end{figure}

Figure~\ref{chan_spec} shows the {\it Chandra} ACIS spectrum of \ax\
folded with the detector response. As recognized from the figure, the
X-ray spectrum is hard; the observed flux was $7\times10^{-15}$ \ergs\
(0.5--2 keV) and $1.3\times10^{-13}$ \ergs\ (2--10 keV), which are
consistent with the previous \asca\ observation (Paper I). In the
spectral fit performed below, we assume Galactic absorption with a
hydrogen column density fixed at 2.6$\times10^{20}$ cm $^{-2}$, as
estimated from HI observations by Dickey \& Lockman (1990).

First we fit the spectrum with a single power law continuum with an
intrinsic
absorption at source frame. As a result, we obtained a photon index of
1.2$\pm$0.2 and a hydrogen column density of $(7\pm 1) \times 10^{22}$
cm$^{-2}$. The fit was not acceptable ($\chi^2$ = 79 for the degree of
freedom of 56), however, leaving a soft excess below $\simeq$ 1.2 keV.
Although we have to bear in mind possible calibration uncertainties in
the current response matrices, we could attribute this to the presence
of a scattered and/or less absorbed component (i.e., partial covering
model). Such components have been observed in many of nearby Seyfert~2
galaxies (e.g., \citet{tur97}). Accordingly, we fit the spectrum with
two absorbed power laws with different absorptions and normalizations,
represented by the formula $A\:E^{-\Gamma} [ f\:e^{-N_{\rm H1} \sigma
(E)} + (1-f)\:e^{-N_{\rm H2} \sigma (E)} ]$, where $E$ is the
source-frame photon energy in keV, $\Gamma$ is the photon index, $A$
is the normalization at 1 keV, $\sigma (E)$ is the photo-electric
absorption cross section (Morisson \& McCammon 1983), $N_{\rm H1}$ and
$N_{\rm H2}$ are hydrogen column densities, and $f$ represents the
covering fraction.  We find that the model can reproduce the observed
continuum well. The best fit parameters are given in
Table~\ref{chan_fits}; the hydrogen column density is $N_{\rm H1}$ =
$(2.3\pm1.1)\times10^{23}$ cm$^{-2}$ and the absorption corrected
source-frame 2--10~keV luminosity goes up to $7.2^{+3.6}_{-2.0} \times
10^{44}$ erg s$^{-1}$ with an intrinsic photon index of $1.93\pm0.44$.

Finally, we add an iron K emission line at the source frame, which is
expected as a result of reprocessing from the X-ray absorbing
matter. We here consider only a narrow line and assume a 1$\sigma$
intrinsic width of 0.02 keV: this corresponds to the velocity width
of $\sim$900 km s$^{-1}$, which is chosen to be comparable to the widths of
optical emission lines (600 km s$^{-1}$; Paper I).
We obtain  the center energy to be
6.51$\pm$0.14 keV with an equivalent width of 148$^{+150}_{-142}$ eV
at the source frame (Table~\ref{chan_fits}). The center energy
indicates that it originates from cold (or warm) ambient matter. The
maximum allowed value for the equivalent width of the iron K line,
$\simeq$300 eV, rules out that the observed spectrum consists purely
of a scattered or reflected component, which would produce an
equivalent width of 1--3 keV (e.g., Matt, Brandt, and Fabian
1996). This gives strong evidence that a significant fraction of the
observed emission comes directly from the central source.  We plot the
best fit model of each component in Figure~\ref{chan_spec} with
residuals in the lower panel.

The relation between the obtained source-frame equivalent width of
iron-K line and the absorption column density is consistent with the
presence of surrounding matter around the central source in a
spherical or torus-like geometry. For example, in the simplest case
that a spherical gas with a uniform density surrounds the central
source, we expect an equivalent width of about 100--300 eV for a
column density of $N_{\rm H} = (1-3)\times 10^{23}$ cm$^{-2}$ assuming
a power law spectrum with $\Gamma = 1.7$ (e.g., \citet{ino85};
\citet{lea93}). For a typical obscuring torus model as defined by
Ghisellini, Haardt and Matt (1994) in their Figure~1, an iron-K
emission is produced with an equivalent width of 40--110 eV for $N_{\rm
H} = (1-3)\times 10^{23}$ cm$^{-2}$ when we view through the torus
having an opening angle $\theta = 30$\degr\ (see their Figure~3). The
source-frame equivalent width of 148$^{+150}_{-142}$ eV is explained
by this torus model having the same column densities as in the line of
sight.

\subsection{$J$-band Spectroscopic Observations with IRCS/Subaru}

\subsubsection{Observation and Data Reduction}

$J$-band spectroscopic observations of \ax\ were done with
InfraRed Camera and Spectrograph (IRCS; \citet{kob00}) attached to
Subaru telescope on 21 March 2001.
A $J$-band grism, which covers a wavelength region from
1.18$\mu$m to 1.38$\mu$m with the
fifth order, and a reflective slit with $0.\!^{\prime\prime}6$
width were used.
In the configuration, the spectral resolution ($R$) was  343 and
the dispersion was 3.7 {\AA} pixel$^{-1}$.
The spatial scale was $0.\!^{\prime\prime}058$ pixel$^{-1}$, which
is over-sampling under the $0.\!^{\prime\prime}35$
seeing condition during the observations.
Because the seeing size was smaller than the slit width,
the higher spectral resolution than $R$ of 343 was achieved for the object.
The spectra were taken at 2 positions, hereafter A and B positions,
of the slit which separate 7$^{\prime\prime}$ apart.
One set of a dithering pattern consists of
four spectra taken at A, B, B, and A positions.
At each position 120 s integration was made and 5 dithering
sets were taken.
Thus, the total effective integration time was 2400 s.
The observation was done
at the object airmass from 1.1 to 1.2.
Before the observation,
a field A0 type star (SAO42433, $V=6.6$ mag)
was also observed at an airmass of
1.2 for the atmospheric absorption correction.

The data were reduced in the following manner;
at first from each set of A and B positions or
B and A positions combinations, the B position image was subtracted
from the A position image.
Most of night sky emission lines were removed with the process, but
for the strongest night sky lines,
they could not be removed completely,
thus we removed the remaining night sky emissions,
using the sky spectra along the slit direction.
After that all A minus B images were
flat fielded with a flat image which was
made from subtracting images taken with
a slit-illumination lamp (halogen lamp)
off from images with the lamp on.
Next, all the flat-fielded images were stacked and
the wavelength calibration of the stacked image was done using
night sky lines in the raw object frame.
The uncertainty of the wavelength calibration was 10 {\AA} (r.m.s.).
After optimum extraction method was applied  to extract one
dimensional spectral data from the two dimensional stacked image,
spectra taken at A position and B position were summed.
Atmospheric absorption and a sensitivity function
were corrected using spectra of the standard star which
were reduced in the same manner.
Flux calibration was done based on the $J$-band magnitude
of the object as described below.

During the telescope acquisition of $J$-band spectroscopic observation,
four images with a 45 s integration time each were also taken with the
$J$-band filter of Mauna Kea Observatory Near-Infrared filter set.
For the flux calibration, three photometric standard stars were observed
(UKIRT Faint Standard stars;
FS021[$J$=12.976mag], FS125[$J$=10.789mag], and FS127[$J$=11.947mag]).
Four images of \ax\ were taken at different position in the
detector, thus at first we subtracted each image from another
image in order to subtract sky background.
After that these sky subtracted images were flat-fielded with
a flat image which was made from subtracting images taken with
the field-illumination lamp (halogen lamp)
off from images with the lamp on.
In each flat-fielded image,
the signal to noise ratio of the object is high,
thus we measured the count rate in  each image, independently.
The measured count rates were consistent within 6 \%.
The standard star frames were reduced in the same manner.
Count rates of each star taken at different positions of the detector
were consistent within 2 \%.
Based on the scatter of calculated count rate to magnitude conversion factor
for the 3 standard stars,
the uncertainty of the flux calibration was estimated to be 0.05 mag.
The measured $J$-band magnitude of AX~J08494+4454 is $17.79\pm0.08$ mag,
and the flux calibration of the $J$-band spectrum was done with
the $J$-band magnitude.
The size of the object in the $J$-band image was FWHM of 5.9 pixel
($0.\!^{\prime\prime}35$), which is the same as those of other stellar
objects.
No extended component was detected.
Because the slit width was wider than the
object size, all parts  of the object should be sampled.

\subsubsection{Results}

\begin{figure}
\plotone{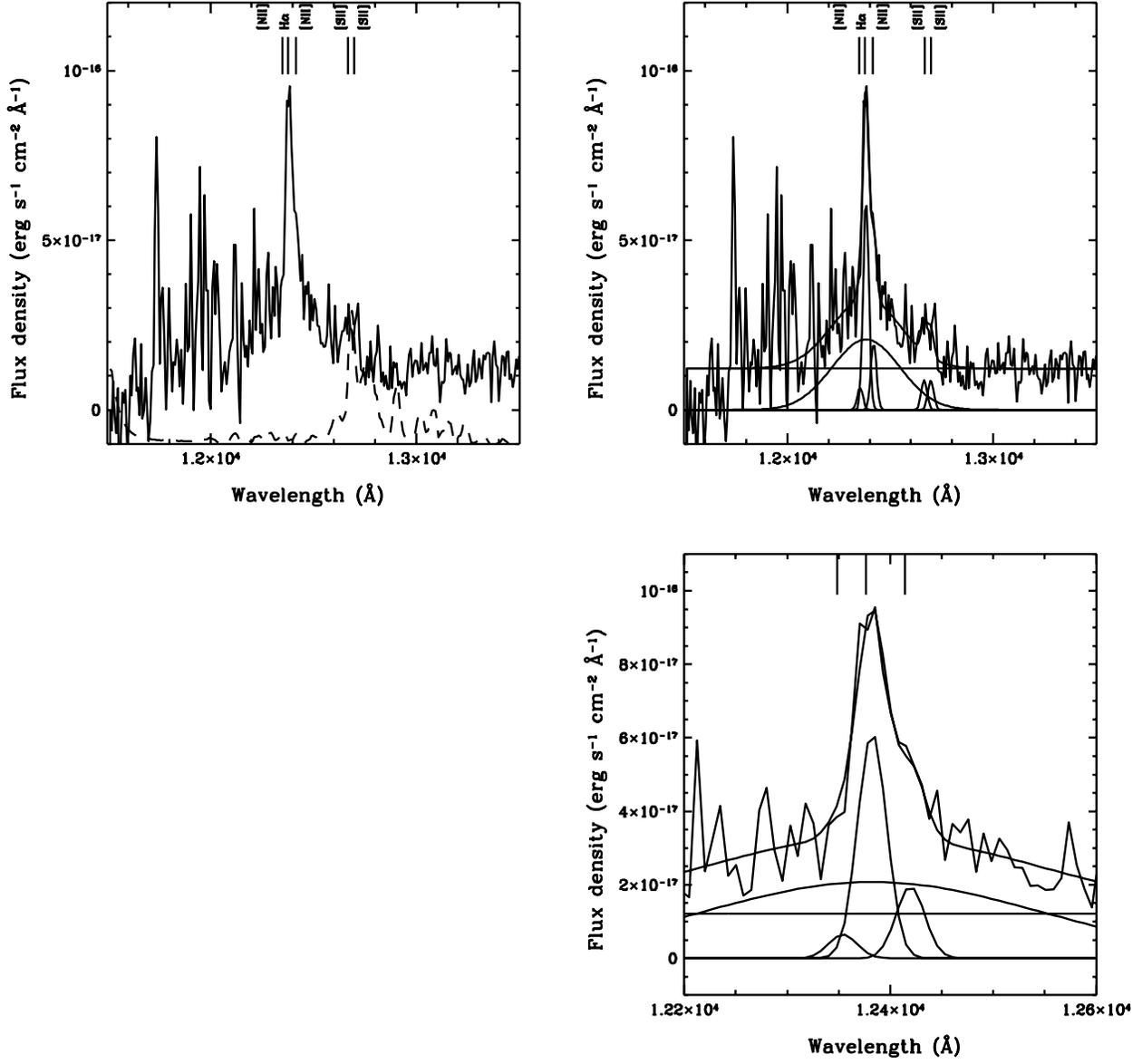}
\caption{$J$-band spectrum of AX~J08494+4454,
a) (upper left) with a night sky spectrum in arbitrary unit (dashed line),
and
b) (upper right)  with the best fit model spectrum.
c) (lower right) same as b), but zoomed in the H$\alpha$, \nii\ wavelength
region.
Identifications of emission lines are shown  in each panel.
\label{AL12_spec}}
\end{figure}

The $J$-band spectrum of the object is shown in Figure~\ref{AL12_spec}.
A strong narrow H$\alpha$ emission line is detected, and it
shows a shoulder in the red side, which comes from \niib\ emission line.
A continuum emission of the object is also detected.
There is a weak bump at the wavelength corresponding to \sii\ emission
lines. It should be noted that the \sii\ region suffers 
from a strong night sky emission line.
The narrow-H$\alpha$ line strength and the continuum flux level
are consistent with those obtained in the previous 
$J$-band spectrum of \ax\ (Paper II).
Although in the previous spectrum no significant 
broad-H$\alpha$ emission line was detected,
in the continuum emission of the new $J$-band spectrum
there is a bump in the wavelength range from 12000{\AA} to 13000{\AA}, 
and an existence of a broad H$\alpha$ emission line is suggested. 
Because of a low efficiency of the grism in the wavelength range
below 12300 {\AA}, the signal to noise ratio in the wavelength range
was low, and the blue part of the broad component is uncertain.
The broad component could be an artifact of an error of the 
sensitivity correction. 
In order to examine whether the broad H$\alpha$ emission line
is due to an error of the sensitivity correction, 
we apply the sensitivity function which is
used to the spectrum of \ax\ to a spectrum of
a bright G5V type star taken in the next night of the 
observation of \ax. The corrected spectrum of the
star is consistent with the model spectrum of 
a G5V type star \citep{kur79} within 10\% 
in the wavelength range above 12000{\AA}, 
and we expect that the broad H$\alpha$ feature is 
not due to an error of the sensitivity correction.
It should be noted that
the probable broad emission line has a much larger velocity width
($\sim9000$ km s$^{-1}$) than that assumed in Figure~1 of Paper II
(3000 km s$^{-1}$).

In order to determine the strength and the velocity width of these
narrow and broad emission lines, fitting to the obtained spectrum
with the $\chi^{2}$ minimization method is applied
in the wavelength range from 12300{\AA} to 13500{\AA}.
In the fitting, we assume that all the redshifts for
the narrow-H$\alpha$, broad-H$\alpha$, narrow-\nii, and
narrow-\sii\ emissions are the same, and the velocity widths of
the narrow-emission lines are also the same.
The flux ratio between \niia\ and \niib\ is fixed.
The best-fit wavelengths, fluxes, and velocity widths of
each emission lines
are summarized in Table~\ref{Tab_emi} and the result of the fitting is
shown in Figures 1b and 1c.
 One sigma uncertainties of
each free parameters are listed also. All of the uncertainties of
emission line fluxes are dominated by $J$-band photometry
uncertainty, except for the flux of \sii\ emission lines.

The resulting redshift of the narrow-H$\alpha$ emission line is
$0.8866\pm0.0008$, which agrees with  that obtained in the
optical spectrum ($z=0.8858\pm0.0002$; PaperI) and obtained
in the previous $J$-band spectrum ($z=0.8857\pm0.0002$; Paper II).
The spectral resolution measured with night sky emission lines in the
frame is FWHM of 30{\AA}, which corresponds to 8 pixel and
$0.\!^{\prime\prime}47$ in the spatial direction.
Resolution for the object emission
line should be higher than that of the night sky lines
because the FWHM of the object ($0.\!^{\prime\prime}35$)
is smaller than the slit width in the spatial direction.
Considering the FWHM difference between the object and the slit
images, we estimated the resolution for the object to be 22 {\AA},
which corresponds to 533 km s$^{-1}$. Using this value as
instrumental broadening, we estimated intrinsic velocity widths
of narrow- and broad-emission lines to be FWHM of $590\pm60$ and
$9400\pm1000$ km s$^{-1}$, respectively.

The probable existence of the strong broad H$\alpha$ emission line 
and non-detection of a broad H$\beta$ emission line suggest 
\ax\ is similar to Seyfert 1.9 galaxies.
However, the width of the broad Balmer emission line corresponds to
the smallest value of high-luminosity radio-quiet QSOs
($M_V < -28$ mag), and is the largest value of low-luminosity
radio-quiet QSOs ($M_V > -25$ mag) \citep{mci99}.
Interpolating the two samples of radio-quiet QSOs, we expect that
the velocity width of \ax\ is consistent with QSOs with
$M_V$ of about $-26$ mag. Therefore, the large velocity width
of the broad-line of \ax\ suggests that the object has
a high-luminosity QSO nucleus in its center.

The narrow-line width is
consistent with that of narrow-\oiii\ emission line in
the optical wavelength
(600 km s$^{-1}$; Paper I) and that of typical values
of narrow-line region of Seyfert galaxies \citep{whi92}.
The \niib\ to narrow-H$\alpha$ flux ratio corresponds to the
smallest value among Seyfert 2 galaxies \citep{vei87}. The large \oiii\ to
narrow-H$\beta$ flux ratio (Paper I) distinguishes the object
from HII region like objects.

\subsection{Optical Photometric Observation and Results}

Optical photometric observations in Mould $B$, $V$, $R$, and $I$ bands
were made with Tektronix 2048$\times$2048 CCD camera
at University of Hawaii (UH) 88$^{\prime\prime}$ telescope
on 1999 March 5 and 6. Exposure times of the observations
were 3600, 2100, 1800, and 1200 s in $B$, $V$, $R$, and $I$ bands,
respectively. During the observing run, 50 Landolt's standard
stars \citep{lan92} were observed.
The seeing size was $1.\!^{\prime\prime}0$
and the pixel scale was $0.\!^{\prime\prime}22$ pixel$^{-1}$.
The imaging data were analyzed with IRAF.
Bias subtraction and flat-fielding with dome flats were performed.
Based on the scatters of the count rate-to-magnitude conversion
factors of the observed Landolt's standard stars,
the uncertainties in the photometric calibrations were estimated to
be 0.03, 0.06, 0.03, and  0.06 mag in $B$, $V$, $R$, and $I$ bands,
respectively.
The $B$, $V$, $R$, and $I$ band magnitudes were measured to be
$22.14\pm0.06$, $21.40\pm0.08$, $20.87\pm0.06$,
and $20.20\pm0.08$ mag, respectively.
The uncertainty includes the
flux calibration uncertainty as described above,
uncertainty of flat-fielding in the object frame, and
the sky subtractions.
These results are summarized in Table~\ref{Tab_mag}.
The $I$ band
magnitude is consistent with a previous measurement
($20.3\pm0.3$; Paper II).
The $I$ band magnitude of the western object is $19.81\pm0.08$ mag and
the summed magnitude of the eastern and the western objects
is $19.24$ mag. This is $1.1$ mag brighter than that
estimated from the optical spectrum (20.3 mag; Paper I).
The discrepancy can be due to an underestimate of the
object brightness in Paper I,
because the slit might not have been centered on the object.
In the $B$-, $V$-, and $R$-band images, \ax\ has
slightly larger image size
(FWHM of 5.1 pixel) than stellar objects in each frame
(FWHM of 4.4 pixel).
In the $I$-band, the image is elongated by a guiding error,
no measurement on the image size was made.

\section{Discussions: Discrepancy Between Optical and X-ray Absorption}

The hard X-ray and narrow-emission line properties of \ax\
are consistent with those of Seyfert 2 galaxies.
Its X-ray spectrum can be fit with an
absorbed power law with a column density of $N_{\rm H} \simeq
(2.3\pm1.1) \times 10^{23}$ cm$^{-2}$ at the source frame.
The results on the X-ray spectrum generally agree with the picture of
the unified scheme for type-2 AGNs \citep{awa91}, except that
the intrinsic 2--10 keV luminosity ($7.2^{+3.6}_{-2.0}
\times10^{44}$ ergs s$^{-1}$) is
larger than those of nearby Seyfert 2 galaxies.
The hard X-ray to \oiii\ luminosity ratio of
the \ax\ follows the extension of those of Seyfert 1 and 2 galaxies
(\citet{hal98}).

The probable existence of the strong H$\alpha$ broad-emission line
conflicts with the large X-ray absorption,
which corresponds to $A_V$ of $130\pm60$ mag with
Galactic extinction curve and gas-to-dust ratio.
In order to quantitatively examine the discrepancy,
we estimated the amount of dust absorption to the
broad-line region with three indicators;
Balmer decrements, hard X-ray to H$\alpha$ luminosity ratio,
and shape of optical continuum emission.

We can only put lower limit
on the Balmer decrement of \ax\,
because broad H$\beta$ emission line was not detected in the
optical spectrum (Paper I).
The upper limit on the flux of the broad H$\beta$ emission line
with the velocity width of 9400 km s$^{-1}$ is
$1.8 \times 10^{-15}$ erg s$^{-1}$ cm$^{-2}$.
Thus, the lower limit on the Balmer decrement to the broad-line
region is 4.8, and we obtain the lower limit on the absorption to
the nucleus as $A_V$ of 1.2 mag, assuming that an intrinsic Balmer
decrement is the typical value in broad-line QSOs (3.1;
\citet{bro01}), and the nuclear absorption follows the
Galactic extinction curve.

Based on the relation between hard X-ray and H$\alpha$ luminosity,
the luminosity of the broad H$\alpha$ emission of object with
hard X-ray luminosity of $7.2^{+3.6}_{-2.0} \times 10^{44}$
erg s$^{-1}$ cm$^{-2}$, should be $(0.5 - 40) \times 10^{43}$
erg s$^{-1}$ cm$^{-2}$ \citep{war88}.
The measured H$\alpha$ flux corresponds to
H$\alpha$ luminosity of $(3.9\pm0.5)
\times 10^{43}$ erg s$^{-1}$ cm$^{-2}$,
which falls within the above range. The upper limit on the
absorption to the H$\alpha$ emission is estimated to be
$A_{\alpha}$ of 2.7 mag, that is converted to $A_V$ of 3.4 mag with the
Galactic extinction curve.

\begin{figure}
\plotone{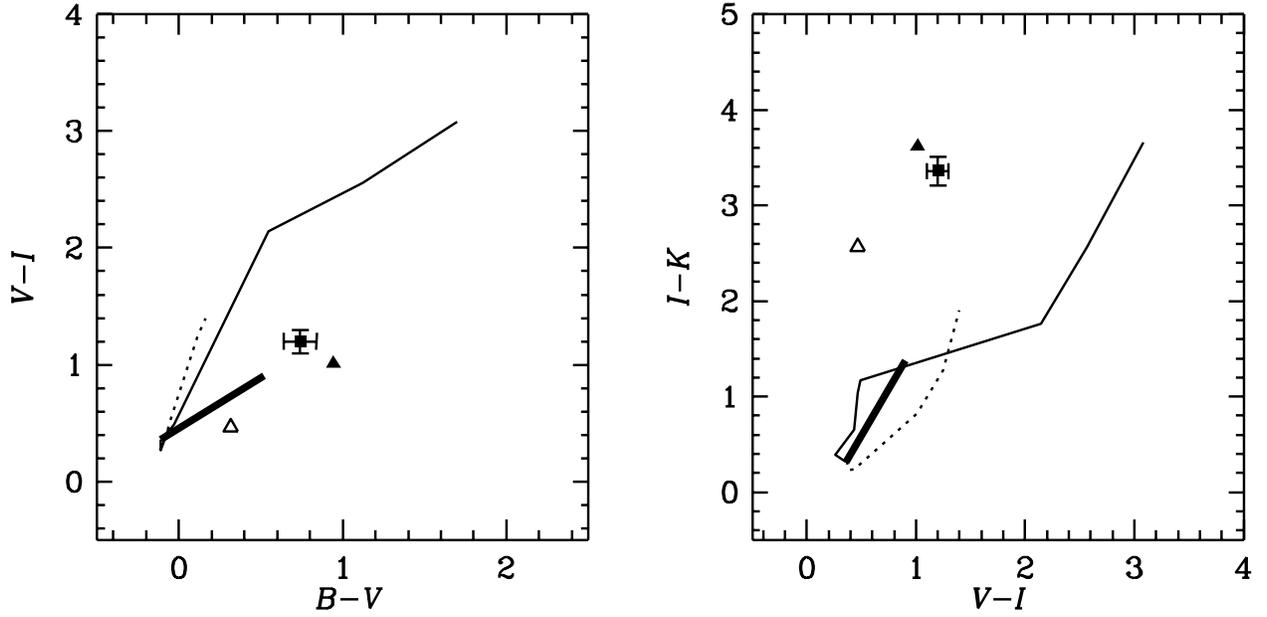}
\caption{
$B-V$ vs. $V-I$ (left) and $V-I$ vs. $I-K$ (right)
diagrams for  \ax. Filled squares represent colors of
\ax. Open and filled triangles indicate colors of average
QSO spectrum with no absorption and absorption with $A_V$ of 1 mag.
Thin solid and dashed lines represent colors of elliptical and
disk galaxy models at redshift of 0.8858 with ages from 0.01 Gyr to 5
Gyr. 
Thick solid lines indicate effect of reddening on young stellar
population model with an absorption of $A_V = 1$ mag.
\label{AL12_color}}
\end{figure}

The reddening to the nucleus is also estimated from the shape of optical
and near infra-red continuum emission, if the optical continuum is
dominated by the reddened nuclear emission.
\ax\ has a red $I-K$ color and, there are three possibilities to the
origin of the red color: reddened nuclear continuum, old stellar
population of host galaxy, and reddened young stellar population.
In Figure~\ref{AL12_color}, $B-V$, $V-I$, and $I-K$ colors
of \ax\ are plotted together with colors of an average QSO spectrum
 \citep{bro01},
those  with reddening of $A_V = 1$
mag, and colors of model galaxies.
We do  not use  the $R$- and $J$-band magnitudes, because the
wavelength ranges are affected by strong [O~II]$\lambda$3727 and
H$\alpha$ emission lines, respectively.
The average QSO colors have scatter of about 0.2 mag (r.m.s.)
\citep{ric01}.
For galaxy models, two galaxy color evolution models are
plotted (Kodama, Arimoto 1997):
1) Elliptical model (plotted with a solid line)
in which star formation occurs during the first 0.353 Gyr with
an initial mass function with a slope of 1.20;
after that the galaxy evolves passively.
The model parameters well reproduce the reddest and
brightest ($M_V=-23$ mag) class elliptical galaxy in the Coma cluster
(Kodama et al. 1998).
2)
Disk model (plotted with a dotted line) is used
in which star formation occurs constantly
with the same initial-mass function as that in the elliptical model.
The colors of these models at redshift of 0.9
with ages from 0.01 Gyr to 5 Gyr,
which corresponds to the age of the universe at the redshift,
are shown with tracks.
The colors of the reddened ($A_V\sim1$ mag) average QSO spectrum
(filled triangles)  are
close to the observed colors of \ax.
While \ax\ does not have as red color as old elliptical model
in $V-I$ color, which reflects the size of 4000{\AA} break at the redshift
of
0.9, thus an old stellar population can not be the origin of
the optical continuum.
Considering the direction of reddening vector, we can also reject
the possibility that an absorbed young stellar population makes
the red color of \ax.
Therefore, the reddened nuclear continuum is the most plausible
origin of the red optical to near infrared colors.
The $J$-band spectrum of \ax\ also supports the model, because
the source frame equivalent width
of the broad H$\alpha$ line (380$\pm50$ {\AA}) is similar to those of
broad-line QSOs ($300 \pm 100$ {\AA}, \citet{bro01}), and
contamination from stellar continuum emission should be negligible
in the $J$-band. On the other hand,
the optical images of \ax\ are slightly extended.
From deconvolution of an $I$-band image of \ax\, the contribution
of the host galaxy to the total light is estimated to be $45-55$\%
(Paper II). 
There are $0.2-0.3$ mag discrepancy between the colors of 
\ax\ and the $A_V\sim1$ mag absorbed
QSO model, and the colors of \ax\ are shifted to
the tracks of the galaxy models.
The discrepancy can be due to the contamination of the host galaxy
component.

\begin{figure}
\plotone{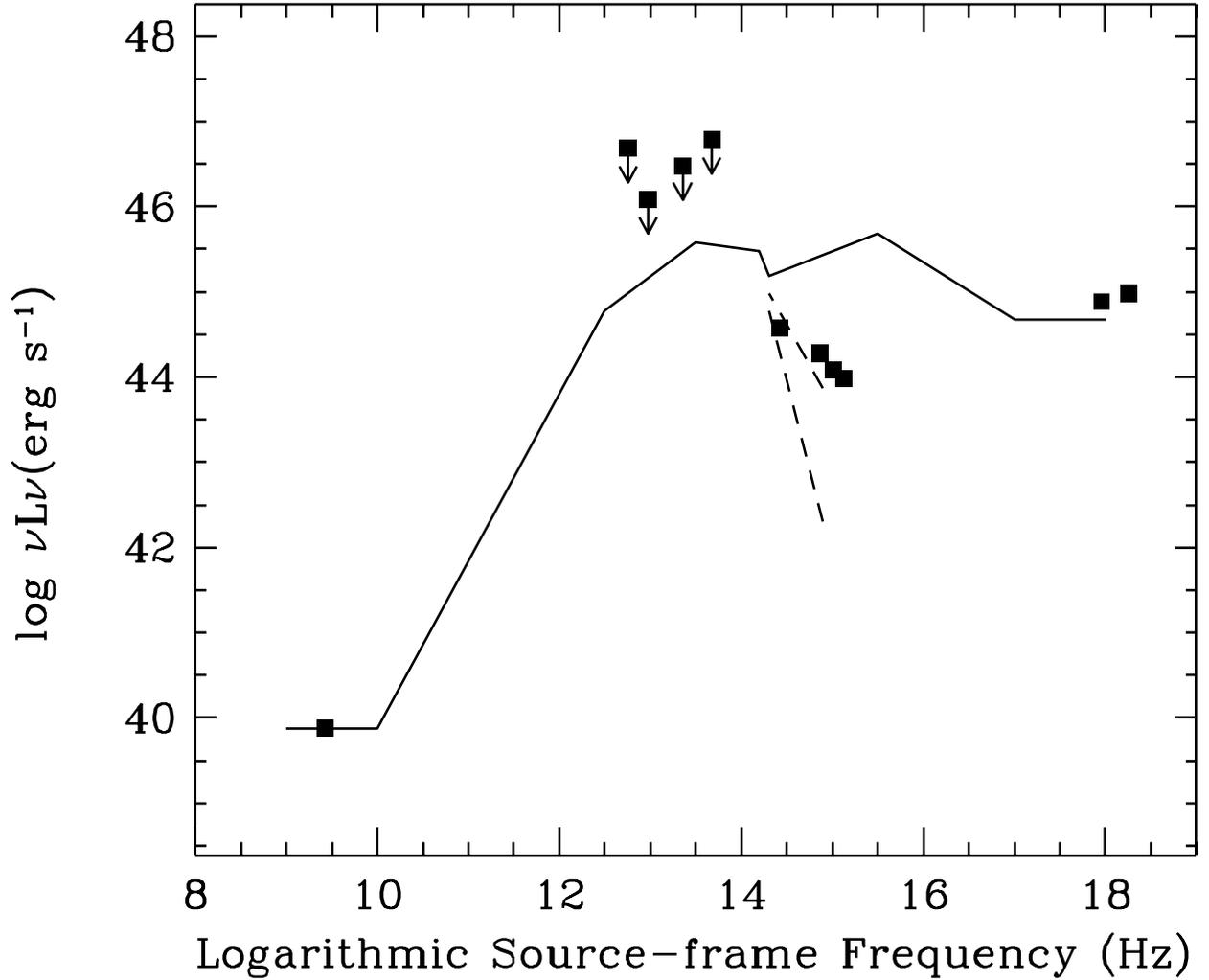}
\caption{
Spectral energy distribution (SED) of AX~J08494+4454.
Filled squares indicate data points of AX~J08494+4454.
Downward arrows represent $IRAS$ upper limits
for AX~J08494+4454 (Paper I). Solid line indicates SED of
an average radio-quiet QSO \citep{elv94}
and is normalized at the data point observed at 1.4 GHz.
Dashed lines represent optical SEDs affected by
dust extinction with $A_V$ of 1 mag (upper) and 2 mag (lower).
\label{AL12_sed}}
\end{figure}

The overall spectral energy distribution (SED) of \ax\ is also consistent
with a  QSO SED affected with $A_V$ of 1 mag in the optical wavelength.
In Figure~\ref{AL12_sed}, the energy densities at 8~keV,
4~keV, $B$-, $V$-, $I$-, $K$-bands (Paper II) and 1.4 GHz \citep{oor87}
are plotted with an average radio-quiet QSO SED taken from
\citet{elv94}. In the hard X-ray range, we plotted the intrinsic 
power-law component determined by the X-ray spectral fitting.
$IRAS$ survey upper limits (Paper I) are also plotted
with downward arrows.
The average QSO SED is normalized to have the
same energy density with \ax\ at 1.4 GHz.
The faint optical magnitudes are
explained well with average QSO spectrum
affected by $A_V$ of 1 mag in the optical wavelength
(upper dashed line in Figure~\ref{AL12_sed}).

To summarize the optical extinction estimations,
the extinction to the nucleus is estimated to 
be $A_V$ of $1-3$ mag from the strength of
broad H$\alpha$ line, and the extinction is consistent with
the optical colors and the overall SED of \ax.
The estimated amount of absorption is significantly smaller than
that expected from the X-ray absorption.
The estimated $A_V/N_{\rm H}$ value of \ax\ is
$(0.3 - 2.5)\times10^{-23}$ mag cm$^{-2}$, and more than
20 times smaller than
those measured in the Galaxy (e.g., \citet{pre95}).
Similar discrepancy of values between absorption from broad Balmer
(or Paschen)-line ratio and from X-ray spectrum is
reported for objects with $N_{\rm H} \ge 10^{22}$ cm$^{-2}$ and
with broad-Balmer and/or Paschen lines (e.g., \citet{mai01a}).
In Seyfert 2 galaxies, the amount of absorption determined
from $L$-band reddening is also much smaller than the expected
absorption derived from the hard X-ray observation (\cite{alo97}).
Similar discrepancy is reported even in the line of sight to
the Galactic center (\citet{mae01}).
The discrepancy can be explained with a smaller dust to gas mass
ratio which may due to dust sublimation in the X-ray absorbing matter,
the size difference between optical and X-ray emitting region, or
different dust size distribution in AGNs (\citet{mai01b}).

In hard-X-ray-selected heavily-absorbed AGNs,
there may be a discrepancy of amount of dust reddening
between Seyfert galaxies and QSOs.
\ax\ has the hardest X-ray spectrum in the X-ray sources
detected in the Lynx field with {\it Chandra}, and it
is confirmed that the object is affected by the heavy X-ray absorption,
but \ax\ probably has broad H$\alpha$ emission as strong as normal QSOs.
In contrast, very hard X-ray sources detected in wider area 
{\it ASCA} surveys are identified with low-redshift 
Seyfert 2 galaxies that
do not show strong broad H$\alpha$ emission line at all
\citep{iwa97,aki98}. Their estimated amounts of X-ray absorption
are similar to that of \ax\, but their H$\alpha$ emission lines
are much more affected by dust extinction than in \ax.
The discrepancy between the high-redshift QSO
and the low-redshift Seyferts
may suggest that X-ray luminous QSOs
have different internal structure from
low-luminosity Seyfert galaxies:
on average, QSOs have smaller $A_V/N_{\rm H}$ value than
low-luminous Seyfert galaxies, and
QSOs are less affected by dust reddening.

\acknowledgments

We are grateful to members of UH88$^{\prime\prime}$ telescope,
Subaru Telescope and IRCS team, especially, Drs. Hiroshi Terada, and
Naoto Kobayashi, who gave us invaluable advices on the $J$-band
observations.
We acknowledge Dr. Leisa Townsley for her kind
support on the CTI correction
of the {\it Chandra} ACIS data and Dr. Yoshitomo Maeda for his
critical advice on the analysis.
We would like to thank the referee, Dr. Paul Green, for valuable 
comments on the work.

\clearpage

\clearpage

\clearpage

\clearpage

\clearpage

\begin{deluxetable}{ll}
\tablecaption{Fitting Results of the {\it Chandra}
Spectrum\label{chan_fits}}
\tablewidth{14cm}
\tablehead{\colhead{} &  \colhead{}}
\startdata
$A$ (10$^{-5} $cm$^{-2}$ s$^{-1}$ keV$^{-1}$) & $20^{+36}_{-13}$ \nl
$\Gamma$ & 1.93$\pm$0.44 \nl
$f$ &0.84$^{+0.08}_{-0.13}$\nl
$N_{\rm H1}$ (10$^{22}$ cm$^{-2}$) &23$\pm$11 \nl
$N_{\rm H2}$ (10$^{22}$ cm$^{-2}$) &3.8$\pm$1.9 \nl
Luminosity\tablenotemark{a} (10$^{44}$ erg s$^{-1}$) &7.2$^{+3.6}_{-2.0}$\nl
\tablevspace{0.2cm}
Line Energy (keV)& 6.51$\pm$0.14 \nl
Equivalent Width\tablenotemark{b} (eV) & 148$^{+150}_{-142}$ \nl
\tablevspace{0.2cm}
$\chi^2$/dof & 52.0/53 \nl
\tablenotetext{a}{
The 2--10 keV luminosity of the power law component in the source frame
corrected for absorption.
}
\tablenotetext{b}{Source frame.}
\tablecomments{
The source-frame spectrum (at z=0.8858) is modeled by the formula
$A\:E^{-\Gamma} [ f\:e^{-N_{\rm H1} \sigma (E)} + (1-f)\:e^{-N_{\rm
H2} \sigma (E)} ]$, where $E$ is the source-frame photon energy in keV,
$\Gamma$ is the photon index, $A$ is the normalization at 1 keV,
$\sigma (E)$ is the photo-electric absorption cross section (Morisson
\& McCammon 1983), $f$ is the covering fraction, and $N_{\rm H1}$ and
$N_{\rm H2}$ are hydrogen column densities. Spectral fit was performed
assuming Galactic absorption of $N_{\rm H} = 2.6\times10^{20}$
cm$^{-2}$.
Errors are 90\% confidence limits for a single parameter.
}
\enddata
\end{deluxetable}

\clearpage

\begin{deluxetable}{lllll}
\tablecaption{$J$ band Emission Line Properties. \label{Tab_emi}}
\tablewidth{0pt}
\tablehead{
\multicolumn{1}{c}{Name} &
\multicolumn{1}{c}{Wavelength} &
\multicolumn{1}{c}{Flux} &
\multicolumn{1}{c}{Width} &
\multicolumn{1}{c}{Luminosity} \\
\multicolumn{1}{c}{} &
\multicolumn{1}{c}{({\AA})} &
\multicolumn{1}{c}{(erg s$^{-1}$ cm$^{-2}$)} &
\multicolumn{1}{c}{(km s$^{-1}$)} &
\multicolumn{1}{c}{(erg s$^{-1}$)}
}
\startdata
$[$NII$]$6548     & $12354$     & $2.3       \times 10^{-16}$ &  795 &
$1.0\times10^{42}$ \\
Broad-H$\alpha$   & $12382$     & $8.6\pm1.1 \times 10^{-15}$ &
$9393\pm1016$, 9400\tablenotemark{a} &
$3.9\times10^{43}$ \\
Narrow-H$\alpha$  & $12382\pm5$ & $2.2\pm0.3 \times 10^{-15}$ & $795\pm61$,
590\tablenotemark{a} &
$9.9\times10^{42}$ \\
$[$NII$]$6583     & $12419$     & $6.9\pm1.3 \times 10^{-16}$ &  795 &
$3.1\times10^{42}$ \\
$[$SII$]$6713     & $12664$     & $3.2\pm1.2 \times 10^{-16}$ &  795 &
$1.4\times10^{42}$ \\
$[$SII$]$6730     & $12697$     & $3.1\pm1.2 \times 10^{-16}$ &  795 &
$1.4\times10^{42}$
\enddata
\tablecomments{All values with uncertainty are independent
parameters in the fitting.}
\tablenotetext{a}{After deconvolution of instrumental profile.}
\end{deluxetable}

\begin{deluxetable}{cccccc}
\tablecaption{Results of Optical Photometry. \label{Tab_mag}}
\tablewidth{0pt}
\tablehead{
\colhead{$B$} &
\colhead{$V$} &
\colhead{$R$} &
\colhead{$I$} &
\colhead{$J$} &
\colhead{$K$} \\
\colhead{(mag)} &
\colhead{(mag)} &
\colhead{(mag)} &
\colhead{(mag)} &
\colhead{(mag)} &
\colhead{(mag)}
}
\startdata
$22.14\pm0.06$ & $21.40\pm0.08$ & $20.87\pm0.06$ &
$20.20\pm0.08$ & $17.79\pm0.08$ & $16.9\pm0.1$\tablenotemark{a} \\
 \enddata
\tablenotetext{a}{Taken from Paper II.}
\end{deluxetable}

\end{document}